\documentclass[aps, preprint, showpacs]{revtex4}  
\usepackage{amsfonts}
\usepackage{amsmath}
\usepackage{amssymb}
\usepackage{graphicx}%
\providecommand{\U}[1]{\protect\rule{.1in}{.1in}}

\begin{document}

\title{The environmental dependence of redox energetics of PuO$_{2}$ and $%
\alpha $-Pu$_{2}$O$_{3}$: A quantitative solution from DFT+$U$
calculations}
\author{Bo Sun$^{*}$}
\affiliation{Institute of Applied Physics and Computational
Mathematics, Beijing 100094, P.R. China}
\author{Haifeng Liu}
\affiliation{Institute of Applied Physics and Computational
Mathematics, Beijing 100094, P.R. China}
\author{Haifeng Song}
\affiliation{Institute of Applied Physics and Computational
Mathematics, Beijing 100094, P.R. China}
\author{Guang-Cai Zhang}
\affiliation{Institute of Applied Physics and Computational
Mathematics, Beijing 100094, P.R. China}
\author{Hui Zheng}
\affiliation{Institute of Applied Physics and Computational
Mathematics, Beijing 100094, P.R. China}
\author{Xian-Geng Zhao}
\affiliation{Institute of Applied Physics and Computational
Mathematics, Beijing 100094, P.R. China}
\author{Ping Zhang}
\thanks{Corresponding authors. Email addresses: sun\_bo@iapcm.ac.cn (B. Sun), zhang\_ping@iapcm.ac.cn (P. Zhang)}
\affiliation{Institute of Applied Physics and Computational
Mathematics, Beijing 100094, P.R. China}

\begin{abstract}
We report a comprehensive density functional theory (DFT) + $U$
study of the energetics of charged and neutral oxygen defects in
both PuO$_{2}$ and $\alpha$-Pu$_{2}$O$_{3}$, and present a
quantitative determination of the equilibrium compositions of
reduced PuO$_{2}$ (PuO$_{2-x}$) as functions of environmental
temperature and partial pressure of oxygen, which shows fairly
agreement with corresponding high-temperature experiments. Under
ambient conditions, the endothermic reduction of PuO$_{2}$ to
$\alpha$-Pu$_{2}$O$_{3} $ is found to be facilitated by accompanying
volume expansion of PuO$_{2-x}$ and the possible migration of
O-vacancy, whereas further reduction of $\alpha $-Pu$_{2}$O$_{3}$ is
predicted to be much more difficult. In contrast to the endothermic
oxidation of PuO$_{2}$,\ the oxidation of $\alpha$-Pu$_{2} $O$_{3}$
is a stable exothermic process.
\end{abstract}
\maketitle

\section{Introduction}

Plutonium oxides have long been of great interest not only for their
high importance in nuclear fuel cycle and in long-term storage of
Pu-based radioactive waste, but also their fancy physical properties
and complex chemical reactions. Most of the rich behaviors of Pu
ions in different oxidation states can be attributed to the complex
character of Pu-5$f$ orbitals\cite{ref-1,ref-2,ref-3}, which locates
in the boundary of localized and delocalized 5$f$ states among the
actinides. Since different oxidation states have widely different
solubilities and demonstrate distinct effects on chemical corrosion
of plutonium surface, the redox properties of Pu-oxides play a
crucial role in the plutonium science and the reactor safety.

When exposed to oxygen-rich air, metallic Pu surface oxidizes
readily to a protective dioxide (PuO$_{2}$)
layer\cite{Has2000,ref-4,ref-6}, with a thin layer of the
sesquioxide\ (Pu$_{2}$O$_{3}$) existing in between. Although, under
special aqueous condition, the hydration reaction on PuO$_{2}$
surfaces (PuO$_{2}+x$H$_{2}$O$\rightarrow $PuO$_{2+x}+x$H$_{2}$) has
been
reported to generate higher compound PuO$_{2+x}$ (x$\leq $0.27)\cite{ref-5}%
, it has been theoretically proved to be strongly
endothermic\cite{ref-8} and the products (PuO$_{2+x}$) are
experimentally found to be chemically unstable at elevated
temperatures\cite{ref-7}. Thus, PuO$_{2}$ is the highest stable
oxide as the compound of choice for the long-term storage of Pu.
However, under oxygen-lean conditions (in the vacuum or inert gas),
the PuO$_{2}$-layer can be reduced to sesquioxides
(Pu$_{2}$O$_{3}$), which can promote the hydrogenation of Pu
metal\cite{Has2000,jnm2011}. Besides the
well-known sesquioxide $\beta $-Pu$_{2}$O$_{3}$ in hexagonal structure ($P%
\overline{3}m1$), the low-temperature sesquioxide phase is cubic $\alpha $-Pu%
$_{2}$O$_{3}$ ($Ia\overline{3}$), with ideal stoichiometric
structure similar to the 2$\times $2$\times $2 fluorite PuO$_{2}$\
($Fm\overline{3}m$) supercell containing 25\% O vacancy located in
the 16$c$ ($0.25,0.25,0.25$) sites [see Fig. 1(top)]. In fact the
cubic sesquioxide was detected in an
abnormal body centered cubic PuO$_{1.52}$ form\cite{alpha-1}, since $\alpha $%
-Pu$_{2}$O$_{3}$ is stable only below $600$ K and usually, a mixture
of cubic PuO$_{1.52}$ and PuO$_{1.98}$ can be obtained by partial
reduction of PuO$_{2}$ at high temperature and cooling to room
temperature\cite{alpha-2}.
Meanwhile, the complex Pu-O phase diagram\cite{pha-dia} shows that the PuO$%
_{2}$-Pu$_{2}$O$_{3}$ region mainly consists of widely
nonstoichiometric compounds PuO$_{2-x}$ and has not been fully
described since the nonstoichiometry can remarkably vary the
properties of Pu oxides at high temperatures.

As the essential component of the uranium-plutonium mixed oxide
(MOX) fuels in fast reactors, considerable efforts have been
dedicated to the defect chemistry of Pu
oxides\cite{def-1,def-2,def-3,def-4}, and theoretical models have
been proposed based on the measured properties/parameters such as
the formation energy $E^{f}$ of oxygen defects. More recently,
Minamoto $et$ $al.$\cite{def-5} employed the DFT calculated $E^{f}$
of single neutral O-vacancy to discuss the phase equilibrium of
PuO$_{2-x}$-Pu$_{2}$O$_{3}$, however, the conventional DFT that
applies the local density approximation (LDA) or generalized
gradient approximation (GGA) underestimates the strong correlation
of the 5$f$ electrons and, consequently, describes PuO$_{2}$ as
incorrect ferromagnetic FM conductor \cite{ref-9} instead of
antiferromagnetic AFM Mott insulator reported by
experiment\cite{ref-10}. Fortunately, several beyond-DFT approaches,
the DFT (LDA/GGA)+ $U$\cite{ref-12}, the self-interaction corrected
LDA\cite{ref-2,ref-13}, the hybrid density functional of (Heyd,
Scuseria, and Enzerhof) HSE\cite{ref-14}, and LDA + dynamical
mean-field theory (DMFT)\cite{ref-15} have been developed to correct
the failures of conventional DFT in calculations of Pu oxides. The
effective modification of pure DFT by LDA/GGA+$U$ formalisms has
been confirmed widely in studies of
PuO$_{2}$\cite{ref-16,ref-17,ref-18,ref-19,ref-20,ref-21,ref-22,ref-23,ref-24},
which have reported the AFM-insulator ground state of PuO$_{2}$. \

In this work, we present a comprehensive DFT+$U$ study of the charge
states and formation energies of oxygen point defects in both PuO$_{2}$ and $\alpha $-Pu$_{2}$O$%
_{3}$, and the equilibrium compositions of PuO$_{2-x}$ as functions
of the environmental parameters (temperature and partial pressure of
oxygen).

The rest of this paper is organized as follows. The details of our
calculations are described in Sec. \textbf{2}. In Sec. \textbf{3},
we present and discuss the results. In Sec. \textbf{4}, we summarize
our main conclusions.

\section{DETAILS OF CALCULATION}

The calculations are carried out using the Vienna \textit{ab initio}
simulation package\cite{Vasp} in the framework of the
projector-augmented wave (PAW) method\cite{Blo}. Following our
previous DFT+$U$
studies\cite{ref-16,ref-17,ref-18,ref-19,ref-20,ref-21} of plutonium
oxides, we apply the GGA+$U$ scheme formulated by Dudarev $et$
$al.$\cite{ref-12} to correct the strong on-site Coulomb repulsion
among the Pu-5$f$ electrons. The choice of $U$ and $J$ parameters
are based on an overall agreement between available experimental
data and theoretical results as regarding the basic physical
properties of PuO$_{2}$ and Pu$_{2}$O$_{3}$, including the lattice
constants, bulk modulus, phonon spectra and density of states (DOS),
and the valence electronic DOS. Based on our extensive test
calculations, it is found that the Hubbard $U$ can alter the
electronic-structure properties of PuO$_{2}$ as well as those of
$\alpha $-Pu$_{2}$O$_{3}$. Specifically, when the effective
parameter U$_{eff}$(=$U$-$J$) is less than $2.0$ eV, the calculated
electronic DOS indicates that $\alpha $-Pu$_{2}$O$_{3}$ are metallic
FM-conductor instead of the Mott-insulators. And the combination of
$U$=$4.75$ and $J$=$0.75$ eV is proved to allow the ground-state
properties of both PuO$_{2}$ and $\alpha $-Pu$_{2}$O$_{3}$ to be
reasonably described.  A 2$\times $2$\times $2 supercell derived
from the ideal fluorite PuO$_{2}$ is used, which accommodates $64$ O
and $32$ Pu with the collinear \textbf{1k} AFM order. Following the
recent DFT+$U$ calculations of the ground state of \textbf{1k} AFM
insulating UO$_{2}$\cite{ref-26}, we have carefully tested the use
of such collinear \textbf{1k} AFM order to make the convergence to
the ground state systematically, and strictly abided by the basic
principle that the state with lowest total energy corresponds to the
ground-state. The charged oxygen defects are modeled by adding or
removing electrons from the supercell. The geometry optimization is
carried out until the forces on all atoms are less than $0.01$
eV/\r{A}. The plane-wave cut-off energy of $400$ eV and a 7$\times
$7$\times $7 Monkhorst-Pack\cite{Monk} $k$-point mesh (64
irreducible $k$ points) are applied, and the convergences with
respect to these input parameters are checked to be within $0.001$
eV.

The formation energy of an isolated O-vacancy $E_{v}^{f}$ or interstitial $%
E_{i}^{f}$\ in charge state $q$ can be defined as

\begin{equation}
E_{v}^{f}=E_{vac}-E_{per}+\mu
_{\mathrm{O}}(T,p_{\mathrm{O_{2}}})+qE_{F}, \label{e1}
\end{equation}%
\begin{equation}
E_{i}^{f}=E_{int}-E_{per}-\mu
_{\mathrm{O}}(T,p_{\mathrm{O_{2}}})+qE_{F}, \label{e2}
\end{equation}%

where $E_{vac}$, $E_{int}$, and $E_{per}$ represent the total
energies of the system with an O-vacancy, an O-interstitial, and the
perfect system, respectively. $E_{F}$ is the Fermi energy relative
to the valence band maximum (VBM). $\mu
_{\mathrm{O}}(T,p_{\mathrm{O_{2}}})$ is the oxygen chemical
potential under partial pressure $p_{\mathrm{O_{2}}}$. Since the
available experiments have just provided the related data [in Fig.
4(b)] under the conditions of such high temperature and particularly
low pressure, where the interaction between the gaseous molecules is
negligibly small, the ideal gas approximation of gaseous O$_{2}$ is
expected to be reasonable. Thus, we can use the well-known
thermodynamic expression\cite{abAT-2}

\begin{equation}
\mu _{\mathrm{O}}(T,p_{\mathrm{O_{2}}})=\frac{1}{2}\left( E_{%
\mathrm{O_{2}}}+\tilde{\mu}_{\mathrm{O_{2}}}(T,p^{0})+k_{_{B}}T\ln
(p_{\mathrm{O_{2}}}/p^{0})\right) , \label{e3}
\end{equation}%

where $E_{\mathrm{O_{2}}}$ is the total energy of the oxygen
molecule. For the standard pressure $p^{0}=1$ atm, the values of
 $\tilde{\mu}_{\mathrm{O_{2}}}(T,p^{0})$
have been tabulated in Ref. \cite{chem-potential}. With the
\textquotedblleft$ab$ $initio$ atomic
thermodynamics\textquotedblright\ method\cite{abAT-2}, one can see
that the zero-temperature DFT+$U$ calculations just provide input
parameters (total energies) for the thermodynamic formalism of the
redox energetics [equations (1) and (2)], and the environmental
dependence of $E_{v}^{f}$ and $E_{i}^{f}$\ is therefore dominated by
$\mu _{\mathrm{O}}$, which includes three required and correlated
factors, namely, $T$, $p_{\mathrm{O_{2}}}$ and
$\tilde{\mu}_{\mathrm{O_{2}}}$. When compared to the experimental
value of $2.56$ eV\cite{expt-bind}, the DFT-GGA calculation is known
to overestimate the binding energy of molecular O$_{2}$ by $0.8$
eV/0.5O$_{2}$\cite{binding,ref-16}, and the correction term of $0.8$
eV is
thus applied in this work. The calculated oxidation reaction energies in Pu$%
_{2}$O$_{3}$+$\frac{1}{2}$O$_{2}\rightarrow $ PuO$_{2}$ are $-3.81$ and $%
-3.19$ eV for $\beta $-Pu$_{2}$O$_{3}$ and $\alpha
$-Pu$_{2}$O$_{3}$, respectively, which indicate that $\alpha
$-Pu$_{2}$O$_{3}$ is more stable at $0$ K. \

\section{RESULTS AND DISCUSSION}

\begin{figure}[ptb]
\begin{center}
\includegraphics[width=0.9\linewidth]{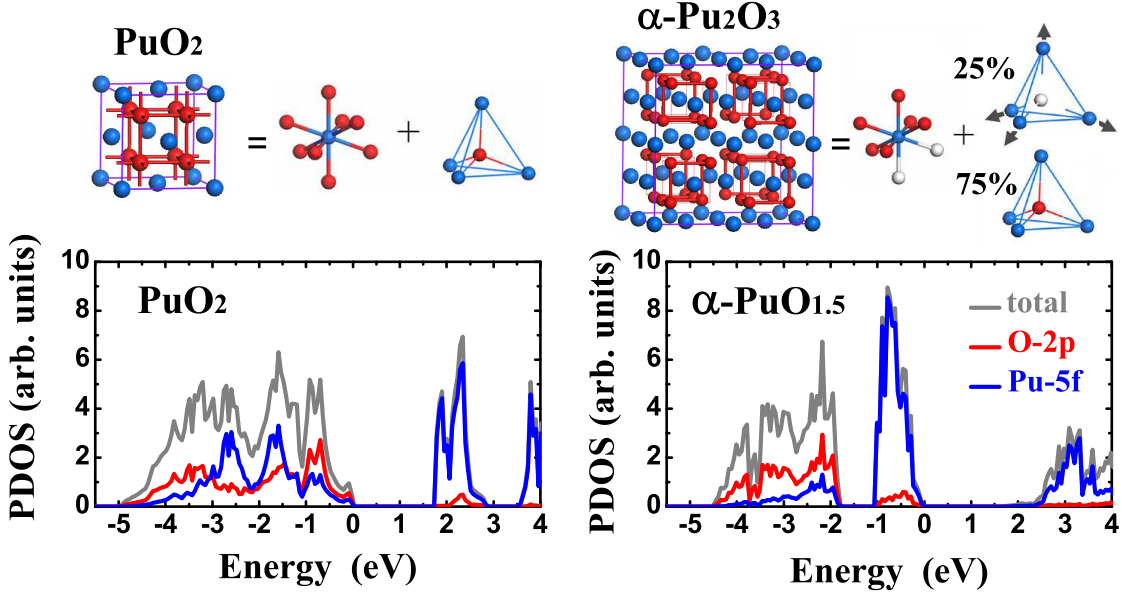}
\end{center}
\caption{(Color online) Top: crystal structures of PuO$_{2}$ and $\protect%
\alpha $-Pu$_{2}$O$_{3}$ and their building blocks with grey arrows
indicate the directions of the atomic displacements. The blue, red
and white spheres denote the Pu, O atoms and O vacancies. Bottom:
the atom-projected density
of states (PDOSs) for Pu-5$f$ and O-2$p$ electrons in PuO$_{2}$ (left) and $%
\protect\alpha $-Pu$_{2}$O$_{3}$ (right). The Fermi level is set to
0 eV. }
\end{figure}

We begin by an overall comparison of the atomic- and
electronic-structure properties between PuO$_{2}$ and $\alpha
$-Pu$_{2}$O$_{3}$ presented in Fig. 1. In PuO$_{2}$, each Pu cation
is bonded to eight O anions and every O is embedded in a regular
Pu-cation tetrahedron, whereas after an isostructure reduction to
$\alpha $-Pu$_{2}$O$_{3}$, the coordination of Pu-cation reduces to
six and O vacancies form in 25\% Pu-tetrahedrons. The calculated
lattice parameter $a_{0}^{\alpha-\mathrm{Pu_{2}O_{3}}}$=$1.120$ nm\
is
close to the experimental values ranging from $1.103$ to $1.107$ nm\cite%
{alpha-1,alpha-2,alpha-3}, only a little larger than
$2a_{0}^{\mathrm{PuO_{2}}} $ of $1.093$ nm (in theory) and $1.079$
nm (in experiment\cite{ref-5}), then an interesting volume expansion
(7.6\% in theory vs 7.4\% in experiment) appears during the
PuO$_{2}\rightarrow \alpha $-Pu$_{2}$O$_{3}$ isostructure reduction.
Moreover, in PuO$_{2-x}$, the abnormal volume expansion is found to
mount up almost linearly with the increasing nonstoichiometry $x$
[see Fig. 4(a) below] in good agreement with experiments. The
orbital-resolved atom-projected density of states (PDOS) suggests
that the reason for such an abnormal issue is the localization of
Pu-5$f$ electrons after the formation
of O-vacancy. Specifically, the highest occupied band (HOB) with a range of $%
-5$ to $0$ eV is mainly the 5$f$(Pu)-2$p$(O) hybridization in
PuO$_{2}$. Whereas in $\alpha $-Pu$_{2}$O$_{3}$, the width of HOB
decreases to $1.2$ eV and the Pu-5$f$ and O-2$p$ occupied bands are
well separated. The deduced
populations of 5$f$ state are $4.1$ and $5.3$ electrons in PuO$_{2}$ and $%
\alpha $-Pu$_{2}$O$_{3}$, respectively. Thus, the formation of O-vacancy in $%
\alpha $-Pu$_{2}$O$_{3}$ or reduced PuO$_{2}$ withdraws 5$f$
electrons from the bonding state and causes the coulomb repulsions
among Pu cations with more localized 5$f$ electrons. Both PuO$_{2}$
and $\alpha $-Pu$_{2}$O$_{3}$
are insulators with the calculated gaps of $1.8$ (the same as experiment\cite%
{ref-6}) and $2.0$ eV.

\begin{figure}[ptb]
\begin{center}
\includegraphics[width=0.85\linewidth]{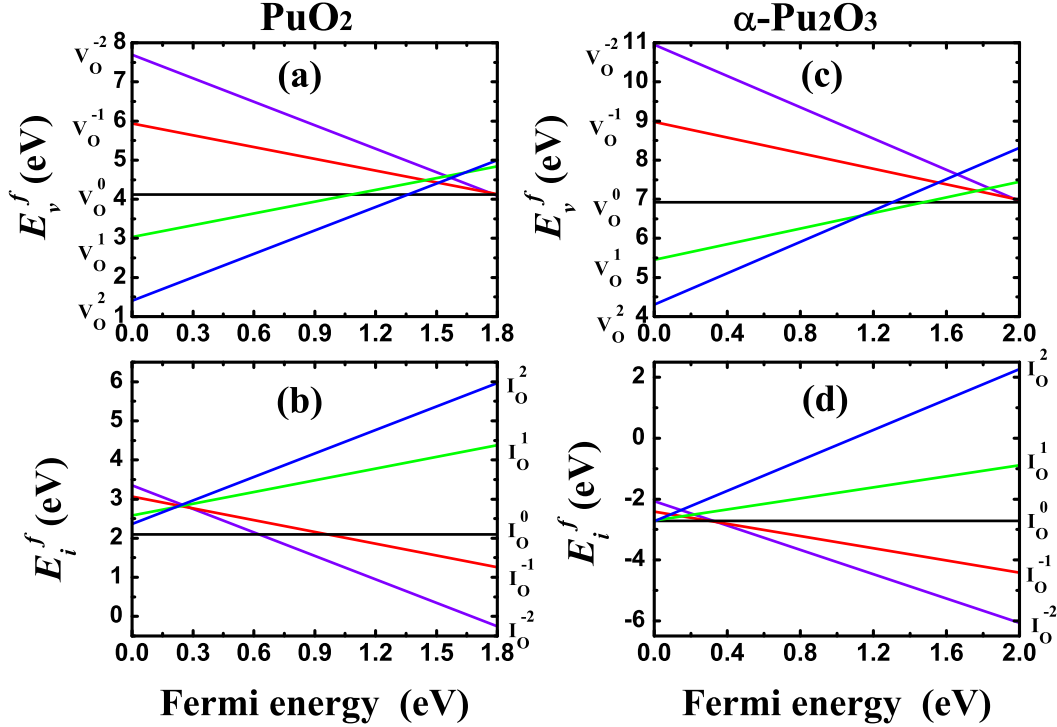}
\end{center}
\caption{(Color online) Formation energy $E_{{}}^{f}$ of oxygen
point defects in different charge states as a function of Fermi
energy in the band gap of PuO$_{2}$ and
$\protect\alpha$-Pu$_{2}$O$_{3}$. }
\end{figure}

For the isolated oxygen defects of different charge states in PuO$_{2}$ and $%
\alpha $-Pu$_{2}$O$_{3}$, the calculated zero-temperature $E_{v}^{f}$ and $%
E_{i}^{f}$\ at the O-rich limit (i.e.
$p_{\mathrm{O_{2}}}$=$p^{0}$=$1$ atm) as functions of the Fermi
energy $E_{F}$ in the insulating band gaps are shown in Fig. 2.
Depending on the location of $E_{F}$, the thermodynamically stable
charge states of O vacancies are found to vary between two values of
$+2$ and $0$ in PuO$_{2}$ [see Fig. 2(a)], and transit from $+2$ to
$0$ via $+1$ in $\alpha $-Pu$_{2}$O$_{3}$ [Fig. 2(c)]. In both
PuO$_{2}$ and $\alpha $-Pu$_{2}$O$_{3}$, it appears that the lowest $%
E_{v}^{f}$ corresponds most of the time to the charge state of $+2$.
And according to the signs and values of $E_{v}^{f}$, one can
conclude that the reduction of PuO$_{2}$ (to $\alpha
$-Pu$_{2}$O$_{3}$) is endothermic, and the further reduction of
$\alpha $-Pu$_{2}$O$_{3}$ will be much more difficult due to the
much higher values of $E_{v}^{f}$. During the reduction of PuO$_{2}$
to nonstoichiometric PuO$_{2-x}$ compounds, the defect-chemistry
study by Stan $et$ $al.$\cite{def-3} predicted that $+2$ O-vacancy
is the dominating defect species at $T=1373$ K in the very low
nonstoichiometry ($x<0.01$) region, namely, $+2$ O-vacancy dominates
only at the beginning of the redox process, then $+1$ O-vacancy in a
limited
intermediate region, and finally neutral O-vacancy in the deep region ($%
x\geq 0.025$). In current calculations based on a 2$\times $2$\times $2 PuO$%
_{2}$ supercell, the nonstoichiometry $x$ is not less than $0.03125$
so that neutral O-vacancy will be the only dominating defect species
according to the foregoing prediction. In both oxides, O
interstitials can be $-2$ or $0$ charged depending on the position
of $E_{F}$. In contrast to the endothermic oxidation of PuO$_{2}$
(to higher oxides PuO$_{2+x}$),\ a minus $E_{i}^{f}$ indicates that
the oxidation of $\alpha $-Pu$_{2}$O$_{3}$ (to PuO$_{2}$) is a
stable exothermic process.

\begin{figure}[ptb]
\begin{center}
\includegraphics[width=0.85\linewidth]{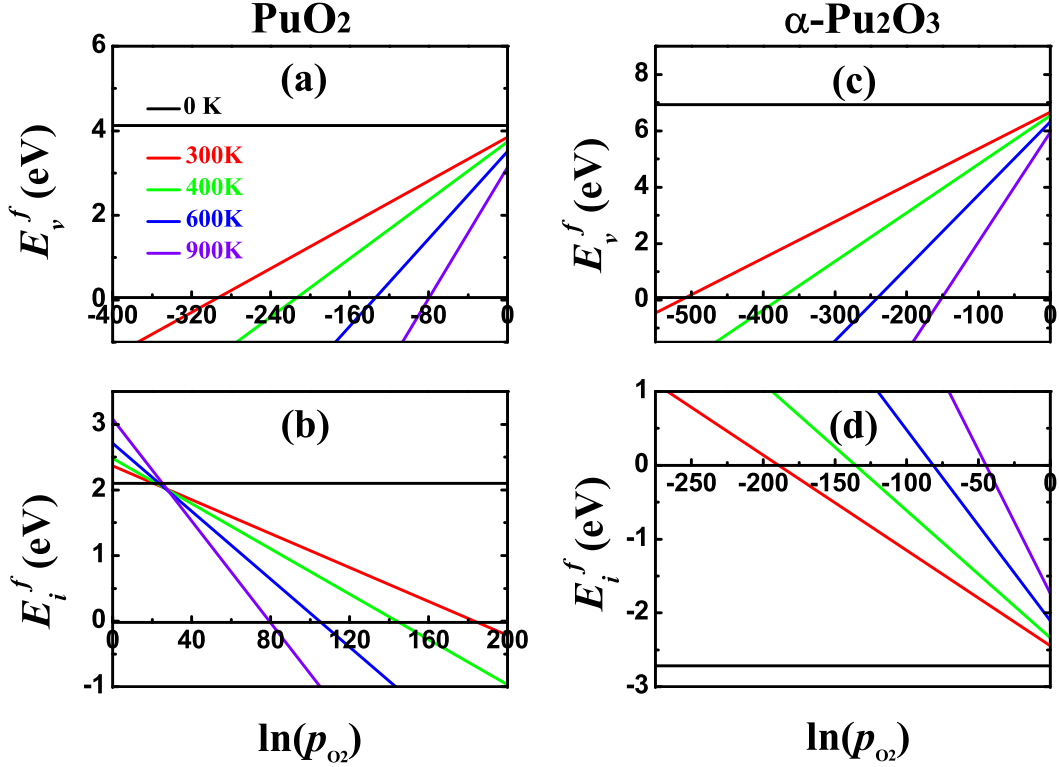}
\end{center}
\caption{ (Color online) Dependence of the calculated formation
energy of oxygen point defects in PuO$_{2}$ and
$\protect\alpha$-Pu$_{2}$O$_{3}$ on temperature $T$ and oxygen
partial pressures $p_{\text{O}_{2}}$.}
\end{figure}

Assuming that both Pu-oxides are in equilibrium with an external
environment and translating $\mu _{\mathrm{O}}$ into the
environmental conditions ($T$ and $p_{\mathrm{O_{2}}}$) according to
Eq. (3), the results presented in Fig. 3(a)-(d) mainly show how a
specific ideal gas approximation corrects the formation energies and
influences the critical oxygen partial pressure
$p_{\mathrm{O_{2}}}^{c}$ with $E_{v}^{f}$=$0$ or $E_{i}^{f}$=$0$ eV,
namely, the evolutions of formation energies\ of neutral O defects
as a function of $T$ and $p_{\mathrm{O_{2}}}$. The O point defects
are found to be sensibly modulated by the environmental $T$ and
$p_{\mathrm{O_{2}}}$ as follows: (i) both $E_{v}^{f}$\ and
$E_{i}^{f}$\ are independent of $p_{\mathrm{O_{2}}}$ when $T=0$ K;
(ii) the $E_{v}^{f}$\ linearly decreases with decreasing
ln$(p_{\mathrm{O_{2}}})$\ when $T>0$ K and reaches zero at a
critical $p_{\mathrm{O_{2}}}^{c}$, which increases with increasing
temperature; (iii) the $E_{i}^{f}$\ linearly decreases with increasing ln$%
(p_{\mathrm{O_{2}}})$\ when $T>0$ K and as $T$ is increased, the
critical $p_{\mathrm{O_{2}}}^{c}$ decreases for PuO$_{2}$, however
increases due to the minus value of $E_{i}^{f}$ for $\alpha $-Pu$_{2}$O$_{3}$%
; (iv) when ln$(p_{O_{2}})$=$0$ (i.e., $p_{\mathrm{O_{2}}}
$=$p^{0}$=$1$ atm), under the ideal-gas approximation the
$\tilde{\mu}_{\mathrm{O_{2}}}(T\neq 0,p^{0})$ term can notably
depress the $E_{v}^{f}$ but increase $E_{i}^{f}$ when compared with
the corresponding formation energy
at $T=0$ K, and it is particularly so at high temperatures. Thus, the $%
\tilde{\mu}_{\mathrm{O_{2}}}(T\neq 0,$ $p^{0})$ term is a
contributory factor to figure out the critical
$p_{\mathrm{O_{2}}}^{c}$ values, since it includes contributions
from rotations and vibrations of molecular O$_{2}$, as well as the
ideal-gas entropy at $1$ atm.

\begin{figure}[tbp]
\begin{center}
\includegraphics[width=0.9\linewidth]{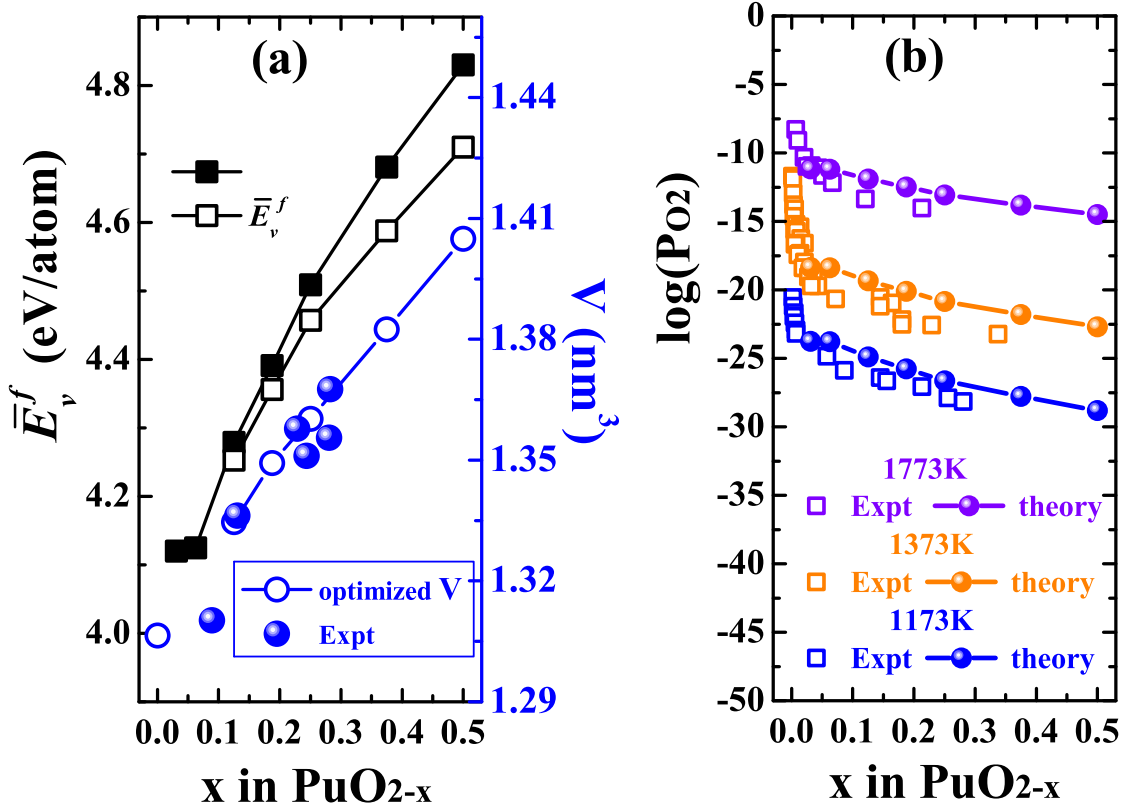}
\end{center}
\caption{ (Color online) (a) The average formation energy
$\bar{E}_{v}^{f}$ of O-vacancy (left axis) and the optimized volume
$V$ of PuO$_{2-x}$ (right axis) varies as the $x$. Solid and open
squares - $\bar{E}_{v}^{f}$ with and without volume optimization;
Blue circles - optimized $V$; Blue spheres - experimental $V$
(compiled from Refs. \cite{alpha-3} and \cite{exp-V}). (b) The
calculated (spheres) and experimental (squares, compiled from Refs.
\cite{exp-V, exp-logP-1, exp-logP-2, exp-logP-3}) relationships of
equilibrium composition of PuO$_{2-x}$ vs temperature and oxygen
partial pressure.  Lines are used to guide the eyes.}
\end{figure}

Now we focus on the environment-dependent equilibrium compositions
of nonstoichiometric PuO$_{2-x}$. The four points of
$p_{\mathrm{O_{2}}}^{c}$\ (at $T$=$300$, $400$, $600$, and $900$ K)
in Fig. 3(a), correspond to four environmental conditions, under
which the PuO$_{1.96875}$ compound is thermodynamically stable. For
the further discussion of other PuO$_{2-x}$ compounds, the averaged
formation energy $\bar{E}_{v}^{f}$ of neutral O-vacancy can be
written as $\bar{E}_{v}^{f}$=$\frac{1}{N_{\mathrm{O-V}}}\left[
E_{vac}-E_{per}+N_{\mathrm{O-V}}\mu _{O}(T,p_{\mathrm{O_{2}}})%
\right] ,$ where $N_{\mathrm{O-V}}$ is the total number of O-vacancy in the Pu$%
_{32}$O$_{64-N_{\mathrm{O-V}}}$ supercell (with $x$=$\frac{N_{\mathrm{O-V}}}{32}$%
). According to the atomic structure of $\alpha $-Pu$_{2}$O$_{3}$
(see Fig.
1), we first adopt the Pu$_{32}$O$_{64-N_{\mathrm{O-V}}}$ supercell with even (%
$N_{\mathrm{O-V}}$= $2$, $4$, $6$, $8$, $12$, and $16$) vacancies
located along the parallel $\langle 111\rangle $ diagonals. In Fig.
4(a), the calculated $\bar{E}_{v}^{f}$ (with and without volume
optimization) and the optimized volume $V$ of
Pu$_{32}$O$_{64-N_{\mathrm{O-V}}}$, together with the compiled
experimental $V$ of PuO$_{2-x}$, are plotted as functions of $x$.
Although the volume expansion is found to effectively depress the $\bar{E}%
_{v}^{f}$ when $x\geq 0.25$, the $\bar{E}_{v}^{f}$ appears to
monotonically increase with $x$ and when $x$=$0.5$ it reaches $4.71$
eV, which is much larger than the reduction energy $E^{r}$ of $3.19$
eV in PuO$_{2}\Rightarrow \alpha
$-Pu$_{2}$O$_{3}+\frac{1}{2}$O$_{2}$. This is due to the special and
fully optimized distribution of O-vacancy in the stable $\alpha $-Pu$_{2}$O$%
_{3}$, specifically, vacancies located along four unequal and
nonintersecting $\langle 111\rangle $ diagonals. We then partially
optimize the O-vacancy distribution based on the
Pu$_{32}$O$_{64-N_{\mathrm{O-V}}}$ supercell with optimized $V$ (for
$N_{\mathrm{O-V}}$= $4$, $6$, $8$, $12$), and the corresponding
$\bar{E}_{v}^{f}$ is found to decrease with the increasing $x$ and
tend to approach $E^{r}$ of $3.19$ eV. To accomplish the optimum
distribution, there may be many complex and correlated mechanisms
and processes, among which the possible migration of O-vacancy and
the accompanying volume expansion of the PuO$_{2-x}$ matrix are
doubtless very important and will be studied in our next work.

Figure 4(b) shows equilibrium compositions of PuO$_{2-x}$ reported
by this work as functions of environmental $T$ and
$p_{\mathrm{O_{2}}}$, which appear to be in good agreement with
experimental data at high temperatures of $1173$ K, $1373$ K and
$1773$ K. Thus, under special high-temperature and low-pressure
conditions, the ideal-gas approximation is a reasonable enough
request for a quantitative determination of the environment ($T$ and
$p_{\mathrm{O_{2}}}$) dependent phase equilibrium of PuO$_{2-x}$. As
a comparison, since the $\tilde{\mu}_{\mathrm{O_{2}}}(T,p^{0})$ term
under the ideal-gas approximation is not considered, the equilibrium
compositions of CeO$_{2-x}$ and HfO$_{2\pm x}$ reported by recent
theoretical studies\cite{APL-2005,APL-2007} are found to be only
qualitative agreement (instead of quantitative) with the
corresponding high-temperature experiments. At low-temperature
conditions, our results (not shown here) indicate that it is very
hard to achieve the desired thermo-equilibrium conditions. However,
with a reducing atmosphere such as CO or H$_{2}$, these conditions
can be generated. Furthermore, in our DFT+$U$ calculations, it is
found that some cluster structures of O vacancies can efficiently
decrease the average formation energy $\bar{E}_{v}^{f}$, the fact of
which indicates that at low temperatures the optimized configuration
of O vacancies in the PuO$_{2-x}$ matrix is the actual key factor
that influences the corresponding thermo-equilibrium conditions,
whereas at high temperatures the O vacancies potentially have an
even distribution.

\section{CONCLUSIONS}

In summary, based on the DFT+$U$ framework and the ideal-gas
approximation, we systematically investigate the environmental
dependence of the redox energetics of PuO$_{2}$ and $\alpha
$-Pu$_{2}$O$_{3}$. Our results clearly reveal that the reduction of
PuO$_{2}$ is an endothermic process, whereas the reduction of $\alpha $-Pu$%
_{2}$O$_{3}$ will be much more difficult due to the higher formation
energies of O-vacancy. In contrast to the endothermic oxidation of
PuO$_{2}$, a minus formation energy of O-interstitial indicates that
the oxidation of $\alpha $-Pu$_{2}$O$_{3}$ is exothermic.
And the reported equilibrium compositions of PuO$_{2-x}$%
\ show quantitative agreement with experiments. Our current study
may provide a guiding line to the understanding of redox behaviors
of plutonium oxides.

\section{ACKNOWLEDGMENTS}
This work was supported by the Foundations for Development of
Science and Technology of China Academy of Engineering Physics under
Grants No. 2010B0301048 and No. 2011A0301016.\


\begin{thebibliography}{99}                                                                                               %


\bibitem{ref-1} S. Y. Savrasov and G. Kotliar, Phys. Rev. Lett. 84 (2000)
3670.

\bibitem{ref-2} L. Petit, A. Svane, Z. Szotek, W. M. Temmerman, Science
301 (2003) 498.

\bibitem{ref-3} K. T. Moore and G. van der Laan, Rev. Mod. Phys. 81
(2009) 235.

\bibitem{Has2000} J. M. Haschke, Los Alamos Science 26 (2000) 253.

\bibitem{ref-4} M. T. Butterfield, T. Durakiewicz, I. D. Prodan, G. E.
Scuseria, E. Guziewicz, J. A. Sordo, K. N. Kudin, R. L. Martin, J.
J. Joyce, A. J. Arko, K. S. Graham, D. P. Moore, and L. A. Morales,
Surf. Sci. 600 (2006) 1637.

\bibitem{ref-6} M. T. Butterfield, T. Durakiewicz, E. Guziewicz, J. Joyce,
A. Arko, K. Graham, D. Moore, and L. Morales, Surf. Sci. 571 (2004)
74.

\bibitem{ref-5} J. M. Haschke, T. H. Allen, and L. A. Morales, Science
287 (2000) 285.

\bibitem{ref-8} P. A. Korzhavyi, L. Vitos, D. A. Andersson and B. Johansson,
Nature Mater. 3 (2004) 225.

\bibitem{ref-7} T. Gouder, A. Seibert, L. Havela, and J. Rebizant, Surf.
Sci. 601 (2007) L77.

\bibitem{jnm2011} L. N. Dinh, J. M. Haschke, C. K. Saw, P. G. Allen, W.
McLean II, J. Nucl. Mater. 408 (2011) 171.

\bibitem{alpha-1} W. H. Zachariasen, Metallurgical Laboratory Report,
CK-1530, 1944; T. D. Chikalla, C. E. McNeilly, R. E. Skavdahl, J.
Nucl. Mater. 12 (1964) 131.

\bibitem{alpha-2} IAEA technical reports series, No. 79, International Atomic Energy Agency, Vienna, 1967.

\bibitem{pha-dia} H. A. Wriedt, Bull. All. Ph. Dia., 11(2) (1990) 184-202.

\bibitem{def-1} T. M. Besmann and T. Lindemer, J. Nucl. Mater. 130 (1985)
489; \textit{ibid}. 137 (1986) 292.

\bibitem{def-2} A. Nakamura, J. Nucl. Mater. 201 (1993) 17.

\bibitem{def-3} M. Stan and P. Cristea, J. Nucl. Mater. 344
(2005) 213.

\bibitem{def-4} C. Gu\'{e}neau, C. Chatillon, and B. Sundman, J. Nucl. Mater.
378 (2008) 257.

\bibitem{def-5} S. Minamoto, M. Kato, and K. Konashi, J. Nucl. Mater. 412 (2011) 338.

\bibitem{ref-9} C. E. Boettger and A. K. Ray, Int. J. Quantum Chem. 90 (2002) 1470.

\bibitem{ref-10} C. McNeilly, J. Nucl. Mater. 11 (1964) 53.

\bibitem{ref-12} S. L. Dudarev, G. A. Botton, S. Y. Savrasov, C. J.
Humphreys, and A. P. Sutton, Phys. Rev. B 57 (1998) 1505.

\bibitem{ref-13} L. Petit, A. Svane, Z. Szotek, W. M. Temmerman, and G. M.
Stocks, Phys. Rev. B 81 (2010) 045108.

\bibitem{ref-14} I. D. Prodan, G. E. Scuseria, J. A. Sordo, K. N. Kudin, and R. L.
Martin, J. Chem. Phys. 123 (2005) 014703.

\bibitem{ref-15} Q. Yin and S. Y. Savrasov, Phys. Rev. Lett. 100 (2008) 225504.

\bibitem{Vasp} G. Kresse and J. Furthm\"{u}ller, Phys. Rev. B 54 (1996)
11169.

\bibitem{Blo} P. E. Bl\"{o}chl, Phys. Rev. B 50 (1994) 17953.

\bibitem{ref-16} B. Sun, P. Zhang, and X.-G. Zhao, J. Chem. Phys. 128 (2008) 084705.

\bibitem{ref-17} B. Sun, and P. Zhang, Chin. Phys. B 18 (2008) 1364.

\bibitem{ref-18} P. Zhang, B.-T. Wang, and X.-G. Zhao, Phys. Rev. B 82 (2010) 144110.

\bibitem{ref-19} H. Shi, M. Chu, and P. Zhang, J. Nucl. Mater. 400 (2010)
151.

\bibitem{ref-20} B. Sun, H. Liu, H. Song, G. Zhang, H. Zheng, X.-G. Zhao, and P. Zhang, J. Nucl. Mater. 426 (2012)
139.

\bibitem{ref-21} H. Shi, and P. Zhang, J. Nucl. Mater. 420 (2012)
159.

\bibitem{ref-22} D. A. Andersson, J. Lezama, B. P. Uberuaga, C. Deo, and S.
D. Conradson, Phys. Rev. B 79 (2009) 024110.

\bibitem{ref-23} G. Jomard, B. Amadon, F. Bottin, and M. Torrent, Phys. Rev.
B 78 (2008) 075125.

\bibitem{ref-24} G. Jomard, and F. Bottin, Phys. Rev.
B 84 (2011) 195469.

\bibitem{ref-26} B. Dorado, B. Amadon, M. Freyss, and M.
Bertolus, Phys. Rev. B 79 (2009) 235125.

\bibitem{Monk} H. J. Monkhorst and J. D. Pack, Phys. Rev. B 13 (1976)
5188.

\bibitem{abAT-2} K. Reuter and M. Scheffler, Phys. Rev. B 65 (2001)
035406.

\bibitem{chem-potential} \textit{NIST-JANAF Thermochemical Tables}, 4th ed.,
edited by J. Chase (American Chemical Society, Washington, DC,
1998).

\bibitem{expt-bind} K. P. Huber and G. Herzberg, \textit{Molecular Spectra
and Molecular Structure IV: Constants of Diatomic Molecules} (Van
Nostrand Reinhold, New York, 1979).

\bibitem{binding} D. A. Andersson, S. I. Simak, B. Johansson, I. A.
Abrikosov, and N. V. Skorodumova, Phys. Rev. B 75 (2007) 035109.

\bibitem{alpha-3} J. C. Boivineau, J. Nucl. Mater. 60 (1976) 31.

\bibitem{APL-2005} Y. Jiang, J. B. Adams, and M. Schilfgaarde, Appl. Phys.
Lett. 87 (2005) 141917.

\bibitem{APL-2007} C. Tang and R. Ramprasad, Appl. Phys. Lett. 91 (2007)
022904.

\bibitem{exp-V} T. L. Markin and M. H. Rand, Thermodynamics, vol. 1 (IAEA,
Vienna, 1966) p. 145.

\bibitem{exp-logP-1} L. M. Atlas and G. J. Schlehman, in: A.E. Kay, M.B. Waldron (Eds.),
Plutonium 1965, Chapman and Hall, London, 1965, p. 838.

\bibitem{exp-logP-2} G. C. Swanson, Los Alamos National Laboratory Report
(LA-6063-T), 1975; R. E. Woodley, J. Nucl. Mater. 96 (1981) 5.

\bibitem{exp-logP-3} O. T. Sorensen, in: Proceedings of the Conference in Baden,
10¨C13 September 1975, North-Holland, Amsterdam, 1976.

\end{thebibliography}
\end{document}